
%
%
%
%
  \font\twelverm=cmr10 scaled 1200       \font\twelvei=cmmi10 scaled 1200
  \font\twelvesy=cmsy10 scaled 1200      \font\twelveex=cmex10 scaled 1200
  \font\twelvebf=cmbx10 scaled 1200      \font\twelvesl=cmsl10 scaled 1200
  \font\twelvett=cmtt10 scaled 1200      \font\twelveit=cmti10 scaled 1200
  \font\twelvemib=cmmib10 scaled 1200
  \font\elevenmib=cmmib10 scaled 1095
  \font\tenmib=cmmib10
  \font\eightmib=cmmib10 scaled 800
  
\font\elevenrm=cmr10 scaled 1095    \font\eleveni=cmmi10 scaled 1095
\font\elevensy=cmsy10 scaled 1095

%
%

\font\seventeeni=cmmi10 scaled \magstep3

\font\seventeensy=cmsy10 scaled \magstep3

\font\seventeenmib=cmmib10 scaled \magstep3

\newfam\cpfam%



\skewchar\eleveni='177   \skewchar\elevensy='60
\skewchar\elevenmib='177  \skewchar\seventeensy='60
\skewchar\seventeenmib='177
\skewchar\seventeeni='177

\newfam\mibfam%


  \skewchar\twelvei='177   \skewchar\twelvesy='60
  \skewchar\twelvemib='177
%
%
\def\twelvepoint{\normalbaselineskip=12.4pt
  \abovedisplayskip 12.4pt plus 3pt minus 9pt
  \belowdisplayskip 12.4pt plus 3pt minus 9pt
  \abovedisplayshortskip 0pt plus 3pt
  \belowdisplayshortskip 7.2pt plus 3pt minus 4pt
  \smallskipamount=3.6pt plus 1.2pt minus 1.2pt
  \medskipamount=7.2pt plus 2.4pt minus 2.4pt
  \bigskipamount=14.4pt plus 4.8pt minus 4.8pt
  \def\rm{\fam0\twelverm}          \def\it{\fam\itfam\twelveit}%
  \def\sl{\fam\slfam\twelvesl}     \def\bf{\fam\bffam\twelvebf}%
  \def\mit{\fam 1}                 \def\cal{\fam 2}%
  \def\tt{\twelvett}%
  \def\mib{\fam\mibfam\twelvemib}%

  \textfont0=\twelverm   \scriptfont0=\tenrm     \scriptscriptfont0=\sevenrm
  \textfont1=\twelvei    \scriptfont1=\teni      \scriptscriptfont1=\seveni
  \textfont2=\twelvesy   \scriptfont2=\tensy     \scriptscriptfont2=\sevensy
  \textfont3=\twelveex   \scriptfont3=\twelveex  \scriptscriptfont3=\twelveex
  \textfont\itfam=\twelveit
  \textfont\slfam=\twelvesl
  \textfont\bffam=\twelvebf
  \textfont\mibfam=\twelvemib       \scriptfont\mibfam=\tenmib
                                             \scriptscriptfont\mibfam=\eightmib

  \def\xrm{\textfont0=\twelverm\scriptfont0=\tenrm
      \scriptscriptfont0=\sevenrm\rm}
\normalbaselines\rm}


\mathchardef\alpha="710B
\mathchardef\beta="710C
\mathchardef\gamma="710D
\mathchardef\delta="710E
\mathchardef\epsilon="710F
\mathchardef\zeta="7110
\mathchardef\eta="7111
\mathchardef\theta="7112
\mathchardef\kappa="7114
\mathchardef\lambda="7115
\mathchardef\mu="7116
\mathchardef\nu="7117
\mathchardef\xi="7118
\mathchardef\pi="7119
\mathchardef\rho="711A
\mathchardef\sigma="711B
\mathchardef\tau="711C
\mathchardef\phi="711E
\mathchardef\chi="711F
\mathchardef\psi="7120
\mathchardef\omega="7121
\mathchardef\varepsilon="7122
\mathchardef\vartheta="7123
\mathchardef\varrho="7125
\mathchardef\varphi="7127

\def\physgreek{
\mathchardef\Gamma="7100
\mathchardef\Delta="7101
\mathchardef\Theta="7102
\mathchardef\Lambda="7103
\mathchardef\Xi="7104
\mathchardef\Pi="7105
\mathchardef\Sigma="7106
\mathchardef\Upsilon="7107
\mathchardef\Phi="7108
\mathchardef\Psi="7109
\mathchardef\Omega="710A}


\def\beginlinemode{\endmode
  \begingroup\parskip=0pt \obeylines\def\\{\par}\def\endmode{\par\endgroup}}
\def\beginparmode{\endmode
  \begingroup \def\endmode{\par\endgroup}}
\let\endmode=\par
{\obeylines\gdef\
{}}
\def\singlespace{\baselineskip=\normalbaselineskip}

\def\oneandaquarterspace{\baselineskip=\normalbaselineskip
  \multiply\baselineskip by 5 \divide\baselineskip by 4}

\def\oneandahalfspace{\baselineskip=\normalbaselineskip
  \multiply\baselineskip by 3 \divide\baselineskip by 2}
\def\doublespace{\baselineskip=\normalbaselineskip \multiply\baselineskip by 2}

\nopagenumbers
\newcount\firstpageno
\firstpageno=2
\headline={\ifnum\pageno<\firstpageno{\hfil}\else{\hfil\elevenrm\folio\hfil}\fi}
\let\rawfootnote=\footnote             
\def\footnote#1#2{{\singlespace\parindent=0pt
\rawfootnote{#1}{#2}}}
\def\raggedcenter{\leftskip=4em plus 12em \rightskip=\leftskip
  \parindent=0pt \parfillskip=0pt \spaceskip=.3333em \xspaceskip=.5em
  \pretolerance=9999 \tolerance=9999
  \hyphenpenalty=9999 \exhyphenpenalty=9999 }
\def\dateline{\rightline{\ifcase\month\or
  January\or February\or March\or April\or May\or June\or
  July\or August\or September\or October\or November\or December\fi
  \space\number\year}}
\def\received{\vskip 3pt plus 0.2fill
 \centerline{\sl (Received\space\ifcase\month\or
  January\or February\or March\or April\or May\or June\or
  July\or August\or September\or October\or November\or December\fi
  \qquad, \number\year)}}


\hsize=6.5truein
\hoffset=0.0truein
\vsize=8.9truein
\voffset=0truein
\hfuzz=0.1pt
\vfuzz=0.1pt
\parskip=\medskipamount
\overfullrule=0pt      



\def\title                     
  {\null\vskip 3pt plus 0.1fill
   \beginlinemode \doublespace \raggedcenter \bf}

\def\author                    
  {\vskip 6pt plus 0.2fill \beginlinemode
   \singlespace \raggedcenter}

\def\affil        
  {\vskip 6pt plus 0.1fill \beginlinemode
   \oneandahalfspace \raggedcenter \it}

\def\abstract                  
  {\vskip 6pt plus 0.3fill \beginparmode
   \doublespace \narrower }

\def\summary                   
  {\vskip 3pt plus 0.3fill \beginparmode
   \doublespace \narrower SUMMARY: }

\def\pacs#1
  {\vskip 3pt plus 0.2fill PACS numbers: #1}

\def\endtitlepage              
  {\endpage                    
   \body}

\def\body                      
  {\beginparmode}              

\def\head#1{                   
  \filbreak\vskip 0.5truein    
  {\immediate\write16{#1}
   \raggedcenter \uppercase{#1}\par}
   \nobreak\vskip 0.25truein\nobreak}

%
%
\def\lefthead#1{                   
  \vskip 0.5truein                 
  {\immediate\write16{#1}
   \leftline  {\uppercase{\bf #1}}\par}
   \nobreak\vskip 0.25truein\nobreak}
%
\def\inlinerefs{
  \gdef\refto##1{ [##1]}                
\gdef\refis##1{\indent\hbox to 0pt{\hss##1.~}} 
\gdef\journal##1, ##2, ##3, 1##4##5##6{ 
    {\sl ##1~}{\bf ##2}, ##3 (1##4##5##6)}}    
\def\keywords#1
  {\vskip 3pt plus 0.2fill Keywords: #1}
\gdef\figis#1{\indent\hbox to 0pt{\hss#1.~}} 

\def\figurecaptions     
  {\head{Figure Captions}    
   \beginparmode
   \interlinepenalty=10000
   \frenchspacing \parindent=0pt \leftskip=1truecm
   \parskip=8pt plus 3pt \everypar{\hangindent=\parindent}}

%
%
\def\refto#1{$^{#1}$}          

\def\references       
  {\head{References}           
   \beginparmode
   \frenchspacing \parindent=0pt \leftskip=1truecm
   \interlinepenalty=10000
   \parskip=8pt plus 3pt \everypar{\hangindent=\parindent}}

\gdef\refis#1{\indent\hbox to 0pt{\hss#1.~}} 

\gdef\journal#1, #2, #3, 1#4#5#6{              
    {\sl #1~}{\bf #2}, #3 (1#4#5#6)}          

\def\refstylenp{               
  \gdef\refto##1{ [##1]}                               
  \gdef\refis##1{\indent\hbox to 0pt{\hss##1)~}}      
  \gdef\journal##1, ##2, ##3, ##4 {                    
     {\sl ##1~}{\bf ##2~}(##3) ##4 }}

\def\refstyleprnp{             
  \gdef\refto##1{ [##1]}                               
  \gdef\refis##1{\indent\hbox to 0pt{\hss##1)~}}      
  \gdef\journal##1, ##2, ##3, 1##4##5##6{              
    {\sl ##1~}{\bf ##2~}(1##4##5##6) ##3}}

\def\prb{\journal Phys. Rev. B, }

\def\prl{\journal Phys. Rev. Lett., }

\def\endreferences{\body}

%
%

\def\endpage                   
  {\vfill\eject}

\def\endpaper                  
  {\endmode\vfill\supereject}

\def\endit
  {\endpaper\end}


\def\ref#1{Ref.[#1]}                   
\def\Ref#1{Ref.[#1]}                   

\def\Equation#1{Equation [#1]}         
\def\Equations#1{Equations [#1]}       
\def\Eq#1{Eq. (#1)}                     
\def\eq#1{Eq. (#1)}                     
\def\Eqs#1{Eqs. (#1)}                   
\def\eqs#1{Eqs. (#1)}                   
\def\frac#1#2{{\textstyle{{\strut #1} \over{\strut #2}}}}

\def\sla{\raise.15ex\hbox{$/$}\kern-.57em}
\def\leaderfill{\leaders\hbox to 1em{\hss.\hss}\hfill}
\def\twiddle{\lower.9ex\rlap{$\kern-.1em\scriptstyle\sim$}}
\def\bigtwiddle{\lower1.ex\rlap{$\sim$}}
\def\gtwid{\mathrel{\raise.3ex\hbox{$>$\kern-.75em\lower1ex\hbox{$\sim$}}}}
\def\ltwid{\mathrel{\raise.3ex\hbox{$<$\kern-.75em\lower1ex\hbox{$\sim$}}}}
\def\square{\kern1pt\vbox{\hrule height 1.2pt\hbox{\vrule width 1.2pt\hskip 3pt
   \vbox{\vskip 6pt}\hskip 3pt\vrule width 0.6pt}\hrule height 0.6pt}\kern1pt}

%

%

%

%
\physgreek
%

\def\dsl{\raise.15ex\hbox{$/$}\kern-.57em\hbox{$\partial$}}
\def\nsl{\raise.15ex\hbox{$/$}\kern-.57em\hbox{$\nabla$}}
\def\gtwid{\,{\raise.3ex\hbox{$>$\kern-.75em\lower1ex\hbox{$\sim$}}}\,}
\def\ltwid{\,{\raise.3ex\hbox{$<$\kern-.75em\lower1ex\hbox{$\sim$}}}\,}
\def\undr{\raise.3ex\hbox{$\sim$\kern-.75em\lower1ex\hbox{$|\vec
x|\to\infty$}}}

\def\[{\left [}
\def\]{\right ]}
\def\({\left (}
\def\){\right )}







\def\and{a^{\phantom\dagger}}

%
\def\id{\raise.72ex\hbox{$-$}\kern-.85em\hbox{$d$}\,}

\catcode`@=11
\newcount\r@fcount \r@fcount=0
\newcount\r@fcurr
\immediate\newwrite\reffile
\newif\ifr@ffile\r@ffilefalse
\def\w@rnwrite#1{\ifr@ffile\immediate\write\reffile{#1}\fi\message{#1}}

\def\writer@f#1>>{}
\def\referencefile{
  \r@ffiletrue\immediate\openout\reffile=\jobname.ref%
  \def\writer@f##1>>{\ifr@ffile\immediate\write\reffile%
    {\noexpand\refis{##1} = \csname r@fnum##1\endcsname = %
     \expandafter\expandafter\expandafter\strip@t\expandafter%
     \meaning\csname r@ftext\csname r@fnum##1\endcsname\endcsname}\fi}%
  \def\strip@t##1>>{}}

\def\citeall#1{\xdef#1##1{#1{\noexpand\cite{##1}}}}
\def\cite#1{\each@rg\citer@nge{#1}}	

\def\each@rg#1#2{{\let\thecsname=#1\expandafter\first@rg#2,\end,}}
\def\first@rg#1,{\thecsname{#1}\apply@rg}	
\def\apply@rg#1,{\ifx\end#1\let\next=\relax
\else,\thecsname{#1}\let\next=\apply@rg\fi\next}

\def\citer@nge#1{\citedor@nge#1-\end-}	
\def\citer@ngeat#1\end-{#1}
\def\citedor@nge#1-#2-{\ifx\end#2\r@featspace#1 
  \else\citel@@p{#1}{#2}\citer@ngeat\fi}	
\def\citel@@p#1#2{\ifnum#1>#2{\errmessage{Reference range #1-#2\space is bad.}%
    \errhelp{If you cite a series of references by the notation M-N, then M and
    N must be integers, and N must be greater than or equal to M.}}\else%
 {\count0=#1\count1=#2\advance\count1
by1\relax\expandafter\r@fcite\the\count0,%
  \loop\advance\count0 by1\relax
    \ifnum\count0<\count1,\expandafter\r@fcite\the\count0,%
  \repeat}\fi}

\def\r@featspace#1#2 {\r@fcite#1#2,}	
\def\r@fcite#1,{\ifuncit@d{#1}
    \newr@f{#1}%
    \expandafter\gdef\csname r@ftext\number\r@fcount\endcsname%
                     {\message{Reference #1 to be supplied.}%
                      \writer@f#1>>#1 to be supplied.\par}%
 \fi%
 \csname r@fnum#1\endcsname}
\def\ifuncit@d#1{\expandafter\ifx\csname r@fnum#1\endcsname\relax}%
\def\newr@f#1{\global\advance\r@fcount by1%
    \expandafter\xdef\csname r@fnum#1\endcsname{\number\r@fcount}}

\let\r@fis=\refis			
\def\refis#1#2#3\par{\ifuncit@d{#1}
   \newr@f{#1}%
   \w@rnwrite{Reference #1=\number\r@fcount\space is not cited up to now.}\fi%
  \expandafter\gdef\csname r@ftext\csname r@fnum#1\endcsname\endcsname%
  {\writer@f#1>>#2#3\par}}

\def\ignoreuncited{
   \def\refis##1##2##3\par{\ifuncit@d{##1}%
     \else\expandafter\gdef\csname r@ftext\csname
r@fnum##1\endcsname\endcsname%
     {\writer@f##1>>##2##3\par}\fi}}

\def\r@ferr{\endreferences\errmessage{I was expecting to see
\noexpand\endreferences before now;  I have inserted it here.}}
\let\r@ferences=\references
\def\references{\r@ferences\def\endmode{\r@ferr\par\endgroup}}

\let\endr@ferences=\endreferences
\def\endreferences{\r@fcurr=0
  {\loop\ifnum\r@fcurr<\r@fcount
    \advance\r@fcurr by 1\relax\expandafter\r@fis\expandafter{\number\r@fcurr}%
    \csname r@ftext\number\r@fcurr\endcsname%
  \repeat}\gdef\r@ferr{}\endr@ferences}


\let\r@fend=\endpaper\gdef\endpaper{\ifr@ffile
\immediate\write16{Cross References written on []\jobname.REF.}\fi\r@fend}

\catcode`@=12

\citeall\refto		
\citeall\ref		%
\citeall\Ref		%

\catcode`@=11
\newcount\tagnumber\tagnumber=0

\immediate\newwrite\eqnfile
\newif\if@qnfile\@qnfilefalse
\def\write@qn#1{}
\def\writenew@qn#1{}
\def\w@rnwrite#1{\write@qn{#1}\message{#1}}
\def\@rrwrite#1{\write@qn{#1}\errmessage{#1}}

\def\taghead#1{\gdef\t@ghead{#1}\global\tagnumber=0}
\def\t@ghead{}

\expandafter\def\csname @qnnum-3\endcsname
  {{\t@ghead\advance\tagnumber by -3\relax\number\tagnumber}}
\expandafter\def\csname @qnnum-2\endcsname
  {{\t@ghead\advance\tagnumber by -2\relax\number\tagnumber}}
\expandafter\def\csname @qnnum-1\endcsname
  {{\t@ghead\advance\tagnumber by -1\relax\number\tagnumber}}
\expandafter\def\csname @qnnum0\endcsname
  {\t@ghead\number\tagnumber}
\expandafter\def\csname @qnnum+1\endcsname
  {{\t@ghead\advance\tagnumber by 1\relax\number\tagnumber}}
\expandafter\def\csname @qnnum+2\endcsname
  {{\t@ghead\advance\tagnumber by 2\relax\number\tagnumber}}
\expandafter\def\csname @qnnum+3\endcsname
  {{\t@ghead\advance\tagnumber by 3\relax\number\tagnumber}}

\def\equationfile{%
  \@qnfiletrue\immediate\openout\eqnfile=\jobname.eqn%
  \def\write@qn##1{\if@qnfile\immediate\write\eqnfile{##1}\fi}
  \def\writenew@qn##1{\if@qnfile\immediate\write\eqnfile
    {\noexpand\tag{##1} = (\t@ghead\number\tagnumber)}\fi}
}

\def\callall#1{\xdef#1##1{#1{\noexpand\call{##1}}}}
\def\call#1{\each@rg\callr@nge{#1}}

\def\each@rg#1#2{{\let\thecsname=#1\expandafter\first@rg#2,\end,}}
\def\first@rg#1,{\thecsname{#1}\apply@rg}
\def\apply@rg#1,{\ifx\end#1\let\next=\relax%
\else,\thecsname{#1}\let\next=\apply@rg\fi\next}

\def\callr@nge#1{\calldor@nge#1-\end-}
\def\callr@ngeat#1\end-{#1}
\def\calldor@nge#1-#2-{\ifx\end#2\@qneatspace#1 %
  \else\calll@@p{#1}{#2}\callr@ngeat\fi}
\def\calll@@p#1#2{\ifnum#1>#2{\@rrwrite{Equation range #1-#2\space is bad.}
\errhelp{If you call a series of equations by the notation M-N, then M and
N must be integers, and N must be greater than or equal to M.}}\else%
 {\count0=#1\count1=#2\advance\count1
 by1\relax\expandafter\@qncall\the\count0,%
  \loop\advance\count0 by1\relax%
    \ifnum\count0<\count1,\expandafter\@qncall\the\count0,%
  \repeat}\fi}

\def\@qneatspace#1#2 {\@qncall#1#2,}
\def\@qncall#1,{\ifunc@lled{#1}{\def\next{#1}\ifx\next\empty\else
  \w@rnwrite{Equation number \noexpand\(>>#1<<) has not been defined yet.}
  >>#1<<\fi}\else\csname @qnnum#1\endcsname\fi}

\let\eqnono=\eqno
\def\eqno(#1){\tag#1}
\def\tag#1$${\eqnono(\displayt@g#1 )$$}

\def\aligntag#1\endaligntag
  $${\gdef\tag##1\\{&(##1 )\cr}\eqalignno{#1\\}$$
  \gdef\tag##1$${\eqnono(\displayt@g##1 )$$}}

\def\eqalignno#1{\displ@y \tabskip\centering
  \halign to\displaywidth{\hfil$\displaystyle{##}$\tabskip\z@skip
    &$\displaystyle{{}##}$\hfil\tabskip\centering
    &\llap{$\displayt@gpar##$}\tabskip\z@skip\crcr
    #1\crcr}}

\def\displayt@gpar(#1){(\displayt@g#1 )}

\def\displayt@g#1 {\rm\ifunc@lled{#1}\global\advance\tagnumber by1
        {\def\next{#1}\ifx\next\empty\else\expandafter
        \xdef\csname @qnnum#1\endcsname{\t@ghead\number\tagnumber}\fi}%
  \writenew@qn{#1}\t@ghead\number\tagnumber\else
        {\edef\next{\t@ghead\number\tagnumber}%
        \expandafter\ifx\csname @qnnum#1\endcsname\next\else
        \w@rnwrite{Equation \noexpand\tag{#1} is a duplicate number.}\fi}%
  \csname @qnnum#1\endcsname\fi}

\def\ifunc@lled#1{\expandafter\ifx\csname @qnnum#1\endcsname\relax}

\let\@qnend=\end\gdef\end{\if@qnfile
\immediate\write16{Equation numbers written on []\jobname.EQN.}\fi\@qnend}

\catcode`@=12
\callall\Equation
\callall\Equations
\callall\Eq
\callall\eq
\callall\Eqs
\callall\eqs


\referencefile
\twelvepoint\doublespace
\endtitlepage
\newtoks\leadline \headline={\hfil}
\lefthead{On the Thermodynamics of Laughlin Liquid Freezing}
\noindent
Anthony Chan and A.H. MacDonald
\bigskip
\noindent
Department of Physics, Indiana University, Bloomington IN
47405

\bigskip

\noindent
{\bf Abstract.}

The competition between liquid and solid
states of strongly correlated electron systems occurs
in a novel way in a strong magnetic field.
The fact that certain Landau level filling factors are
especially favorable for the formation of strongly
correlated liquid states, gives rise to the fractional
quantum Hall effect.  In this article we discuss some
consequences of the existence of incompressible states with
fractionally charged quasiparticle excitations for the
thermodynamics of the liquid-solid transition.

\bigskip

\lefthead{1. Introduction}
\noindent

In the limit of strong magnetic fields electrons in
two-dimensional (2D) systems are confined to the
single-particle states of minimum quantized kinetic energy,
i.e. to states in the lowest Landau level.  In this limit
all states of the system have the same kinetic energy so
that the ground state, and for $k_B T \ll \hbar \omega_c$ also
all thermodynamic properties, are determined entirely by
electron-electron interactions.  In this article we restrict our
attention completely to the strong magnetic field limit where
only the lowest spin-polarized Landau level is relevant.\refto{llmix}
One single-particle state exists in the lowest
Landau level for each magnetic flux quantum piercing the system
and the electron density is usually measured in terms of the
Landau level filling factor $ \nu \equiv n \Phi_0/B =
n h c / e B $ where $n$ is the areal electron density and
$\Phi_0$ is the electron magnetic flux quantum.
The lowest interaction energy state of 2D electrons
is the Wigner\refto{wigner} crystal state
in which electrons are localized at the sites of a triangular
lattice.  At zero field it is the
kinetic-energy cost of localizing electrons which prohibits
the formation of the crystal state except at extremely low
densities.  In a strong magnetic field, liquid and
crystalline states of electrons compete in a different way;
electrons cannot be localized to a
length smaller than the cyclotron orbit radius corresponding
to the quantized lowest kinetic energy, $\ell \equiv (\hbar c
/e B)^{1/2} $.  Only when $\ell$ is much smaller than the
typical distance between electrons, i.e. only when $ \nu \ll 1$
should the crystal state be expected to be the ground state.
For weaker fields (larger filling
factors) the ground state should be expected to be a
liquid\refto{theory,lev}.

The discovery of the fractional quantum Hall
effect\refto{qhe,fqhe,intro} and its subsequent
explanation\refto{laugh}
established the intricate and exotic nature of the
electron liquid states which occur
in the strong magnetic field limit.  More recently data
from transport experiments\refto{trans}, and also from
magneto-optical\refto{magop}
and frequency-dependent conductivity experiments\refto{micro},
has accumulated\refto{solid,reviews} which suggests
the occurrence of the electron solid state.
The experimental results are consistent with theoretical
expectations that liquid states should tend to be favored
near the filling factors where incompressible states occur
and especially for $\nu$ near $1/m$ where $m$ is an odd integer.
It is at these filling factors that the most stable
incompressible liquid states, those first discovered
by Laughlin\refto{laugh}, occur.  Many aspects of the
experiments which seem to indicate a Wigner crystal state
remain poorly understood, as discussed elsewhere in this
volume, and it seems likely that in real samples disorder
must play an important role\refto{zhu} in determining the way in
which Wigner crystallization is manifested.  In this article
we discuss only the ideal limit in which no disorder is
present.  We take the view\refto{zerofield} that in this limit the transition
between fractional Hall liquid states and the Wigner crystal state
is strongly first order so that we can study the thermodynamics
of the phase transition by comparing separate estimates of the
filling factor and temperature dependence of the free energies
of liquid and solid states.  In Section II we discuss the
physics of fractional Hall liquid states for $\nu$ near
$1/m$ and estimates of the temperature and filling factor
dependence of the free energy which result.  In  Section III
we briefly review what is known about the filling factor and
temperature dependence of the free energy of the Wigner crystal
state.  Our considerations are based primarily on the harmonic
approximation for the lattice state.
The filling factor and temperature dependences
of the free energy are very different for the two states and as we
discuss in Section IV these differences suggest the occurrence of
several unusual features in the phase boudary.  In Section V
we briefly summarize our findings.

\lefthead{2. Thermodynamics of the Liquid States}

\noindent
For pairs of electrons restricted to the lowest
Landau level the possible relative motion states may be
labelled by angular momentum $m = 0, 1, \cdots $.  There
is only one relative motion state of the pair for each
value of $m$ and for the fully spin-polarized states of the
electron gas assumed here
electrons can be found only in states of odd relative
angular momentum.  The Hamiltonian of the system can
be completely specified in terms of the interaction energies
of pair of electrons with relative angular momentum $m$,
$V_m$. For $ N_{\Phi} \equiv A B / \Phi_0 = m (N-1) +1$ and any odd
integer $m$ it is possible to show\refto{intro} that there is only one
state in the many-particle Hilbert space for which the amplitude
for finding any pair of electrons in a state of
relative angular momentum less than $m$ is zero.  (Here $N$ is the
number of electrons and $A$ is the area of the 2D system.)
That state is the strongly-correlated electron liquid
state discovered by Laughlin,\refto{laugh,trug} $|\Psi_L \rangle$.
Since electrons
are closer together when they are in a state of lower relative
angular momentum it follows, for sufficiently short-range
{\it repulsive} interactions, that at this particular
magnetic field the Laughlin state will be separated from all other
states in the Hilbert space by an energy gap $\Delta  \sim V_{m-2}$.
Similary the chemical potential at zero temperature will jump
from a value $\sim V_{m}$ to a value $\sim V_{m-2}$ when the
filling factor increases beyond $\nu=1/m$.
The existence of the energy gap and the associated chemical potential
discontinuity leads to the fractional quantum Hall effect.
The chemical potential discontinuity must also lead to anomolies
in the thermodynamic properties of the electron system at low temperatures
for $\nu$ near $1/m$.

For $N_{\phi} = m (N-1) +1 +N_{qh}$ the
many-body states in which relative angular momenta
less than $m$ can be completely avoided can be mapped to the
many-body states for $N_{qh}$ fermion holes in a Landau level with
a degeneracy equal\refto{chpap} to $\tilde N_{\phi} = N+N_{qh}-1$.
The quasiholes
must have fractional charge $e/m$ since $m$ of them are created when one
electron is removed from the sytem at fixed magnetic fields.  When the
fractional quantum Hall effect occurs states in this class
should be separated from higher energy states by a gap
$\sim \Delta$.  Similarly, it can be convincingly argued\refto{hier,jain,mdj}
from several different points of view that the low-energy states for
$N_{\phi} = m (N-1)+ 1 - N_{qe}$ can be mapped to the many-body states
of $N_{qe}$ fractionally charged fermion particles in a Landau level with
degeneracy $\tilde N_{\phi} = N+1-N_{qe}$.
We refer to these excitations with fractional
negative charge as quasielectrons and use the word quasiparticles to
refer to the fractionally charged excitations generically.
Note that when they are considered as fermions the quasiholes see
both electrons and other quasiparticles as sources of effective magnetic
flux with the flux from quasiparticles directed in opposition to that
from the other sources.
In the following we assume that this picture
can be extended to apply to low-lying excitations as well as to the
ground state.  We estimate the thermodynamic
properties of the electron system by assuming that
its elementary excitations are quasiparticle-quasihole pairs.
Since we are interested in filling factors near $\nu=1/m$
and low-temperatures we will assume that the quasiparticles are
sufficiently dilute that we can neglect their mutual
interactions.  This approximation
is adequate for most purposes\refto{compress} and
as we shall see below some interesting non-trivial effects
exist in the distribution functions for these ideal gases
even when quasiparticle interactions are neglected.

We consider the system at a
constant magnetic field.  Dropping terms $\sim 1/N$ we define
$\tilde N \equiv N - N_0$ where $N_0 =  N_{\phi}/m$ so that by total
charge conservation $N_{qe}=N_{qh}+m \tilde N$.  The energies
of the states of the sytem at a given $N$ and hence at a  given $\tilde N$
may be expressed in terms of the number of quasielectrons
and quasiholes:
$$
E(N_{qe},N_{qh}) =\epsilon_0 N_0 + \epsilon_{qe} N_{qe} + \epsilon_{qh}
N_{qh}.
\eqno(chui1)
$$
Here $\epsilon_0$ is the energy per electron of the incompressible
state while $\epsilon_{qe}$ and $\epsilon_{qh}$ are the energies
to make quasiholes and quasiparticles at fixed magnetic field.
For the physically interesting case of Coulomb interactions between the
electrons estimates\refto{thickness} exist for these three parameters:
For $m=5$ and in $e^2 / \ell$ units\refto{macdsmg}
$\epsilon_{0L} \approx -0.3277$,
$\epsilon_{qh} \approx 0.1072 $, and $\epsilon_{qe} \approx -0.076$.
(We use $e^2 / \ell $ units for energies throughout this paper.)
Notice that the quasielectron-quasihole pair creation energy
$\Delta \equiv \epsilon_{qh}+ \epsilon_{qp}  \approx 0.031 $.  This
means for $N_{\phi}= m (N-1) +1 $ the liquid
ground state is separated from excited states in this approximation
by an excitation gap equal\refto{magrot}
to $\Delta$.  At zero temperature the upward discontinuity in
the chemical potential is $m \Delta$ at $N_{\phi} = m (N-1) +1 $.

We evaluate thermodynamic properties at finite temperature in the
grand canonical ensemble.  The grand partition function is
$$
\eqalignno{
Z_G &= \exp ( -\beta (\epsilon_0 - \mu) N_0 ) \sum_{\tilde N,N_{qe},N_{qh}}
\delta_{N_{qe}-N_{qh},m \tilde N}
 \exp [ \beta \mu N_{qe}/m - \beta F^{(0)}(N_{qe},\tilde
N_{\phi},\epsilon_{qe}) ]\cr
& \exp [ -\beta \mu N_{qh}/m - \beta F^{(0)}(N_{qh},\tilde
N_{\phi},\epsilon_{qh}) ],&(chui2)\cr
}
$$
where $\tilde N_{\phi} = N + N_{qh} - N_{qe}$ is the degeneracy of the
quasiparticle and quasihole Landau levels.  We see below that the
dependence of the quasi-Landau-level degeneracies on the number of quasiholes
and quasiparticles which are present on the chemical potential
unusual.  In \Eq{chui2} $F^{(0)}(N,N_{\phi},\epsilon)$ is the finite
temperature free energy for $N$ non-interacting Fermi particles in
a system with a single energy level of energy $\epsilon$ and
degeneracy $N_{\phi}$, which is known exactly\refto{huang}:
$$
F^{(0)}(N,N_{\phi},\epsilon) = N_{\phi}[ \epsilon \nu +
k_B T (\nu \ln (\nu) + ((1-\nu) \ln (1 - \nu ) ] \equiv N_{\phi} f^{(0)}
(\nu, \epsilon)
\eqno(chui3)
$$
where $\nu \equiv N/N_{\phi}$.  To evaluate \Eq{chui2} we use the
$\delta$ function to eliminate the sum over $\tilde N$.  At a given
$T$ and $\mu$ the most
probable values of $N_{qe}$ and $N_{qh}$ can be determined by
finding the extrema of the summand in \Eq{chui2}.
Setting the derivative with respect to $N_{qe}$ to zero gives
$$
\mu = m \mu_{qe} - (m-1) k_B T [\ln (1 - \nu_{qe}) + \ln (1 - \nu_{qh})]
\eqno(dec1)
$$
while setting the derivative with respect to $N_{qh}$ to zero gives
$$
\mu = - m \mu_{qh} - (m-1) k_B T [\ln (1 - \nu_{qe}) + \ln (1 -
\nu_{qh})].
\eqno(dec2)
$$
Here $\nu_{qp} \equiv N_{qp}/\tilde N_{\phi}$ is the quasiparticle
Landau level filling factor.
The second terms on the right hand side of \Eq{dec1} and \Eq{dec2}
come from the derivatives of the quasiparticle Landau level
degenearacies with respect to the quasiparticle and quasihole
numbers.  In these equations we have defined
$$
\mu_{qp} \equiv {d f^{(0)}(\nu_{qp}, \epsilon_{qp}) \over d \nu_{qp}}
= \epsilon_{qp} + k_B T \ln [\nu_{qp}/(1-\nu_{qp})].
\eqno(dec3)
$$
so that the quasiparticle filling factors are related to their
chemical potentials by the usual Fermi distribution function.
Subtracting \Eq{dec1} and \Eq{dec2} we see that $\mu_{qe}=-\mu_{qh}$ so that
, given $\mu$, $\mu_{qe}$ can be determined by requiring
\Eq{dec1} to be satisfied when $\mu_{qh}=-\mu_{qe}$.
The electron filling factor $\nu \equiv N/ N_{\phi}$ can be related to
the quasiparticle filling factors:
$$
\nu = {1 + \nu_{qe} - \nu_{qh} \over m + (m-1) (\nu_{qe} - \nu_{qh} )}.
\eqno(dec4)
$$
We will be interested in comparing the Helmholtz free energy of the liquid
at a particular filling filling factor with that of the solid.  Noting
that fluctuations become negligible in the thermodynamic limit we
find from \Eq{chui2} that
$$
F \equiv -k_B T \ln Z_G + \mu N =  \epsilon_0 N_0 + \tilde N_{\phi}
[f^{(0)}(\epsilon_{qe},\nu_{qe}) + f^{(0)}(\epsilon_{qh},\nu_{qh})]
\eqno(dec5)
$$
so that the free energy per electron is
$$
f \equiv {\epsilon_0 \over m \nu} + (\nu^{-1} - (m-1))
[f^{(0)}(\epsilon_{qe},\nu_{qe}) + f^{(0)}(\epsilon_{qh},\nu_{qh})].
\eqno(dec6)
$$
To determine $\nu_{qe}$ and $\nu_{qh}$ at a given filling
factor and temperature we express the right hand side of \Eq{dec4}
in terms of the quasiparticle energies and
$z_{qe} \equiv \exp[\mu_{qe}/ (k_B T)] = \exp[-\mu_{qh}/ (k_B T) ]$
and solve for its value.  Given $z_{qe}$ we can easily calculate
$\nu_{qe}$, $\nu_{qh}$ and the free energy. Some typical results
are shown in Fig.[1].  The most noteworthy feature in this figure
is related to the entropy,
$$
S = - {\partial F \over \partial T}.
\eqno(dec7)
$$
At $\nu=1/m$ the ground state is non-degenerate and separated
from excited states by the gap $\Delta$.  It follows that the
entropy vanishes like $\exp [ - \Delta/ (2 k_B T) ] $ at
low temperatures so that the free energy, which must
decrease monotonically with temperature, is nearly constant.
For $\nu \ne 1/m$ the ground state is degenerate
as discussed near the beginning of this section, and the
entropy approaches a constant.  For $\nu < 1/m$, the degeneracy
of the many-body ground state is
$$
g = {N+N_{qh} \choose N_{qh}}
\eqno(dec8)
$$
so that the entropy at zero temperature in the thermodynamic limit is
$$
{S \over N k_B } = (1 + \nu^{-1} - m) \ln [1 + \nu^{-1} - m]
		 - (\nu^{-1} - m) \ln [\nu^{-1} - m].
\eqno(dec9)
$$
Similary the zero temperature entropy for $\nu > 1/m$ is
$$
\eqalignno{
{ S \over N k_B } & = (m - \nu^{-1}) \ln [(m - \nu^{-1})^{-1}-1]\cr
&- (1 - 2 (m - \nu^{-1})) \ln [(1 - 2 (m - \nu^{-1}))/(1-(m - \nu^{-1}))].&
(dec10)\cr}
$$
The finite zero-temperature free energy gives rise to
the linear decrease in free energy with increasing
temperatures which is evident at low temperatures in Fig.[1].
In a more realistic model the entropy would drop to zero on a
temperature scale reflecting quasiparticle interactions\refto{qpi};
however these
interactions would not greatly alter the comparisons between
liquid and solid free energies which we make below.

\lefthead{3. Thermodynamics of the Harmonic Electron Solid}

\noindent
In this section we briefly review known results for
the thermodynamics of the harmonic 2D Wigner crystal in the
strong-magnetic-field limit.  In using the harmonic approximation
for the electron crystal we are ignoring the possibility of large
departures of the electrons from their lattice sites.  The
use of this approximation is consistent with our assumption
that the phase-transition is strongly first order since the
harmonic approximation can then still be valid when the solid melts.
We remark that the harmonic approximation is still useful for
estimating thermodynamic properties at low temperatures even
though true long-range-order in the Wigner crystal cannot
exist at finite temperatures.

Adding the Lorentz force to the classical equations of motion it
is easy to demonstate that the harmonic phonon energies in the
strong magnetic field limit are
related to the zero-field energies\refto{fuk} by
$$
\epsilon_{+} (\vec q) =  \hbar \omega_c + \sum_{\lambda} \epsilon_{\lambda}^2
(\vec q) / 2  \hbar \omega_c
\eqno(3a)
$$
and
$$
\epsilon_{-} (\vec q) = [\prod_{\lambda} \epsilon_{\lambda} (\vec q)] / \hbar
\omega_c.
\eqno(3b)
$$
(See for example Ref.[\cite{macdcote}].
Here $\epsilon_{\lambda}$ is a zero-field phonon energy.)
At long wavelenths the zero-field resonance frequencies are a
purely longitudinal plasmon
mode, $ \epsilon_{pl}^2 (q) = 2 \pi \hbar^2 n
e^2 q / m^*$ and a linearly dispersing transverse mode $ \epsilon_{t}
 = c_{t} q$.  It follows that $ \epsilon_{+} (q) =\hbar \omega_c +
 \epsilon_{pl}^2 / 2 \hbar \omega_c$ while $ \epsilon_{-} (q) =
 c_{t} q \epsilon_{pl} (q) / \hbar \omega_c  \sim q^{3/2}$.

At general wavelengths the zero-field phonon energies can be expressed
in the form\refto{maradudin}
$$
\epsilon_{\lambda,\vec q} = \big({ \hbar^2 e^2 (2 \pi n)^{3/2} \over
m^{*} } \big)^{1/2} \tilde \epsilon (\tilde {\bf q})
\eqno(dec11)
$$
where $\epsilon (\tilde {\bf q})$ is a readily calculable pure number,
$\tilde {\bf q} \equiv n^{1/2} \vec q$, and $n$ is the areal density
of the electron system.  It follows that at strong fields
$$
\epsilon_{+}(\vec q) = \hbar \omega_c + {1 \over 2} (e^2/ \ell) \nu^{3/2}
\sum_{\lambda} \tilde \epsilon_{\lambda}^2 (\tilde {\bf q}) \equiv
\hbar \omega_c + (e^2/\ell) \nu^{3/2} \tilde \epsilon_{+} (\tilde
{\bf q})
\eqno(dec12)
$$
and
$$
\epsilon_{-}(\vec q) = (e^2/ \ell) \nu^{3/2}  \prod_{\lambda} \tilde
\epsilon_{\lambda}(\tilde {\bf q}) \equiv (e^2/\ell) \nu^{3/2} \tilde
\epsilon_{-}(\tilde {\bf q})
\eqno(dec13)
$$
Note that the low energy modes are independent of the electron mass
and therefore must correspond to intra-Landau-level excitations
of the system.
For $k_B T \ll \hbar \omega_c$ the mean harmonic-osillator quantum
number is zero for
the high-frequency cyclotron mode so that the internal energy is given
by $E = N \hbar \omega_c /2 + E_{mad} + N (e^2/ \ell) \nu^{3/2} e(t)$
where
$$
e(t) = { 1 \over 2 N} \sum_{\vec q} [ \tilde \epsilon_{+}(\tilde {\bf
q})
+ (2 n( \tilde \epsilon_{-}(\tilde {\bf q})/t) + 1 ) \epsilon_{-} ( \tilde
{\bf q}) ],
\eqno(dec14)
$$
$E_{mad}$ is the classical point lattice energy, $n(x)=(\exp (x) +
1)^{-1}$ and $t \equiv k_B T / (e^2/\ell) \nu^{3/2}$.
Similarly the entropy has contributions only from the intra-Landau-level
phonon modes and depends on temperature only through the dimensionless
paramter $t$; $ S = N k_B s(t)$ where $s(t)$ is readily evaluated
numerically.   In Fig.[2], Fig.[3], and Fig.[4] we plot results
for the specific heat ($C_V = N k_B e'(t)$), the entropy and the
Helmholtz free energy ($F= N \hbar \omega_c/2 + N (e^2/\ell)
(-0.78213 \nu^{1/2} + \nu^{3/2} f(t)$) respectively.  Here
$f(t)=e(t)-t s(t)$ and the second term in the Free energy is the
Madulung energy.  All quantities are evaluated for the triangular
electron lattice which has the lowest\refto{maradudin}
free energy over the temperature range of interest.  In the next
section we use this free energy to construct an estimate
of the shape of the liquid-solid phase boundary near an incompressible
filling factor.  The free-energy of the solid, unlike that of the
liquid, has a smooth dependence on filling factor.  At low temperatures
the vibrational contribution to the free-energy per electron is
$ f(t) \sim - 9 (e^2/\ell) \nu^{3/2} t^{7/3}$;
this power law follows from the $q^{3/2}$ dispersion of the
intra-Landau-level mode and should be compared to the linear temperature
dependence which occurs except at $\nu=1/m$ for the liquid.
Thus finite temperature effects should generally be expected to favor the
liquid over the solid, as is commonly the case.   An exception
occurs for $\nu$ very close to $1/m$ where there are few low
energy excitations for the liquid.

When leading anharmonic corrections are included the ground state
energy of the Wigner crystal is given by\refto{theory}
$$
\epsilon_S  = N \hbar \omega_c /2 + N (e^2/\ell) [-0.78213 \nu^{1/2} + 0.2410
\nu^{3/2}
 + 0.087 \nu^{5/2}]
 \eqno(3c)
$$
In \Eq{3c} the first term in square brackets
is the energy of the classical Wigner
lattice, the second term is the harmonic zero point energy,
and the last term comes from anharmonic corrections.  (The second term
should be compared with the zero-temperature limit of the
vibrational free energy shown in Fig.[4]).  We see that
for $\nu \sim 1/5$, the region of filling factor we focus on below,
anharmonic corrections are $\sim 10^{-3} (e^2/ \ell)$ per particle.
A similar uncertainty exists in estimates of the energy per particle
of the incompressible state.  Below we take the attitude that
the difference between the ground state energies of liquid and
solid states at $\nu = 1/m$ cannot be determined with sufficient
accuracy and may in practice be altered somewhat by changes in the
quantum confinement width of the two-dimensional electron gas.
We will treat this difference as a phenomenological parameter.

\lefthead{4. The Liquid-Solid Phase Boundary}
\noindent

In Fig.[5] we plot the filling factor dependences of the
free-energies for liquid and solid states obtained using the
approximations discussed in the previous two sections.
These curves were obtained assumming that the temperature
dependence of the small anharmonic contribution to the energy
of the solid can be neglected.
The ground state energy of the liquid state has the cusp
responsible for the fractional quantum Hall effect at
$\nu=1/m$.  The ground state energy of the liquid is lower
than that of the solid for a finite range of filling factors
around $\nu=1/m$.  Well away from $\nu=1/m$ the degeneracy of the
liquid ground state causes the free-energy to decrease more
rapidly for the liquid than for the solid.  We have constructed the
phase diagram for our model of the liquid-solid phase
transition by mapping out the filling factors at which the
free-energy differences cross zero.  The results are shown in
Fig.[6].  There is qualitative agreement between these
phase boundaries and those mapped out by various experimental
probes\refto{reviews} of the two-dimensional electron gas system in a strong
magnetic field in this regime of filling factors.
We emphasize that this level of agreement is achieved without
any adjustable parameters.

The temperature scale in Fig.[6]
should be compared with the classical melting temperature of the
Wigner solid which is $ T_M  \sim 6 \times 10^{-2} \nu^{1/2} (e^2 / \ell
k_B)$.  Thus, in our approximation, the solid phase persists to
temperatures well above the classical melting temperature.  This
behaviour seems to be improbable, although experimental
fingerprints of the solid phase do also seem to persist beyond
the classical melting temperature.  The temperature scale
in our figures is strongly sensitive to the quasiparticle-quasihole
creation energy and could be adjusted downward by adjusting
$\Delta$.  Our results are also strongly sensitive to the
ground state energy difference between the liquid and solid states
at $\nu \equiv 1/m$.  This difference is obtained by subtracting two
quantities which have a relative difference of only
several parts per thousand and must be regarded as highly uncertain.
It may even be altered importantly by changes in the effective
interaction between electrons due to differences in the electronic
width of a heterojunction or quantum well between different samples.
For that reason we plot in Fig.[7a-e] a series of phase diagrams
obtained by arbitrarily adjusting the ground state energy of the
liquid at $\nu = 1/m$ so that is passes through our approximate
value of the ground state energy of the solid at this
filling factor.  Members of this sequence of phase diagrams should also
apply qualitatively, with suitable downward adjustments to the
temperature scale, to the shape of the phase boundary for $\nu$ near
$1/m$ at larger values of $m$ since the ground state energy difference
of the two states is expected to cross zero\refto{lev,recentest} with
increasing $m$.  As the ground state energy of the liquid is
increased the width of the filling factor interval where the
liquid is stable at $T=0$ decreases.  At the same time
the $T=0$ entropy of the liquid at the position of the phase
boundary decreases so that the effect of increasing temperature
in expanding the width of the liquid stability interval is diminished.
This is seen in Fig.[7a] and Fig.[7b] where the phase
boundary lines become more vertical.  When the ground state energy
is raised still further the entropy of the solid along
the phase boundary can increase above that of the liquid as
temperature increases so that, in accordance with the
Clausius-Clapeyron equation\refto{plischke}, the width of the liquid stability
interval decreases with temperature.  As we see in Fig.[7c] and
Fig.[7d] the stability interval can decrease to zero so that the solid
becomes stable at $\nu \equiv 1/m$ over some finite temperature
interval before melting again.  Finally when the ground state at
$\nu=1/m$ is raised above that of the the solid, as in Fig.[7e],
the solid will melt at a temperature which is minimized at $\nu=1/m$.
It seems likely that this is typically the situation for $\nu=1/9$
and may often be the case for $\nu=1/7$.  The behavior at
a particular filling factor in a particular system may depend
on the degree of Landau level mixing {\it i.e.}, on the
electron density\refto{recentest}, as well as on geometric
details of particular systems which influence effective
electron-electron interactions.

\lefthead{5. Summary}

In this article we have taken the view that the transition
between Wigner crystal and fluid states in two-dimensions
in a strong magnetic field is strongly first order.  The
thermodynamics of the phase-transition is then dominated
by the same anomoly which is responsible for the fractional
quantum Hall effect; namely the existence of incompressible
states.  When interactions between fractionally charged
quasiparticles are neglected the entropy of the
liquid has a finite contribution at zero-temperature
which vanishes as the incompressible filling factor
is approached.  As a result the shape of the liquid-solid phase boundary
changes qualitatively as the filling factors along the
phase boundary approach the incompressible filling factor.
When the ground state energies
of the Wigner solid and the incompressible liquid are
nearly identical the liquid can freeze and remelt with
increasing temperature.

\noindent

\vfill\eject

\references

\newtoks\leadline \headline={\hfil}
\noindent
\oneandaquarterspace
\hsize=12.0truecm
\hoffset=+0.4truecm
\vsize=18.0truecm
\voffset=-0.7truecm

\refis{recentest} R. Price, P.M. Platzman, and S. He,
Phys. Rev. Lett. {\bf 70}, 339 (1993); X. Zhu, and S.G.
Louie, Phys. Rev. Lett. {\bf 70},335 (1993); Rodney Price,
Xuejun Zhu, P.M. Platzman, and Steven G. Louie,
Phys. Rev. B {\bf 48}, 11473 (1993).

\refis{magop} H. Buhmann {\it et. al.}, Phys. Rev. Lett.
{\bf 65}, 633 (1990); {\it ibid}, {\bf 66}, 926 (1991).

\refis{zhu} See for example Xuejun Zhu, P.B. Littlewood,
and A.J. Millis, preprint (1993) and work cited therein.

\refis{llmix} The magnetic field induced Wigner crystal has
been seen experimentally for Landau level filling
factors $\nu$ smaller than $\sim 1/5$.  For these
filling factors and carrier densities in typical
two-dimensional electron systems mixing of higher Landau
levels extremely weak.  See for example R. Price, P.M. Platzman,
and S. He, Phys. Rev. Lett. {\bf 70}, 339 (1993).

\refis{reviews} See other articles in this volume.  For
a previous review see R.G. Clark in
{\it Low-Dimensional Electronic Systems}, edited by
G. Bauer, F. Kuchar, and H. Heinrich (Springer-Verlag, Berlin, 1992),
p. 239.

\refis{maradudin} L. Bonsall and A.A. Maradudin, Phys. Rev. B
{\bf 15}, 1959 (1977).

\refis{fuk} H. Fukuyama, Solid State Commun. {\bf 17}, 1323 (1975).

\refis{huang} See for example Kerson Huang, {\it Statistical Physics}
(Wiley, New York, 1987) p. 183.

\refis{thickness} These estimates are for an ideal two-dimensional
electron gas where the spread of the subband wavefunction in the
perpendicular direction is much smaller than the distance between
electrons.  In realistic systems all three quantities are typically
reduced by $\sim 30\%$.

\refis{compress} One example of where this approximation can fail even
at zero temperature is in the calculation of the compressibility
for $\nu$ near $1/m$.
See J.P. Eisenstein, L.N. Pfeiffer, and K.W. West, Phys. Rev. Lett.
{\bf 68}, 674 (1992); {\it ibid} preprint (1993).

\refis{zerofield} This view is generally believed to
be appropriate for a two-dimensional electron gas in the
absence of a magnetic field.  For a calculation of the
quantum melting density based on this philosophy see B. Tanatar and
D.M. Ceperley Phys. Rev. B {\bf 39}, 5005 (1989).

\refis{mdj} For a review see M.D. Johnson and G.C. Canright,
J. Physics A, to appear (1994).  See also G.C. Canright and
M.D. Johnson, Phys. Rev. B, to appear (1994).

\refis{micro} F.I.B. Williams, P.A. Wright, R.G. Clark, E.Y. Andrei,
G. Devill, D.C. Glattli, O. Probst, B. Etienne, C. Dorin,
C.T. Foxon, and J.J. Harris, \prl 66, 3285, 1991; R.L. Willett, {\it et.
al.}, Phys. Rev. B {\bf 38}, 7881 (1988).

\refis{trans} H.W. Jiang, R.L. Willett, H.L. Stormer, D.C. Tsui,
L.N. Pfeiffer, and K.W. West, \prl 65, 633, 1990; V.J. Goldman,
M. Santos, M. Shayegan, and J.E. Cunningham, \prl 66, 3285,
1990; Y.P. Li, T. Sajoto, L. W. Engel, D.C. Tsui,
and M. Shayegan, \prl 67, 1930, 1991.

\refis{plischke} Michael Plischke and Birger Bergersen,
{\it Equilibrium Statistical Physics}, (Prentice Hall, New Jersey,
1989).

\refis{magrot} The neutral excitation gap actually differs from
$\Delta$ once quasiparticle-quasihole interactions are taken
into account.  S.M. Girvin, A.H. MacDonald, and P.M. Platzman,
\prb 33, 2481, 1986.

\refis{lev} D. Levesque, J.J. Weis, and A.H. MacDonald,
\prb 30, 1056, 1984.

\refis{macdcote} R. Cote and A.H. MacDonald, \prb 44, 8759, 1991.

\refis{macdsmg} A.H. MacDonald, and S.M. Girvin, \prb 34, 5639, 1986.
The quasiparticle energies here are the neutral
quasiparticle energies in which the quasiparticles are
created at fixed total electron number.

\refis{jain} J.K. Jain, Phys. Rev. Lett. {\bf 63}, 199 (1989); Adv.
Phys. {\bf 41}, 7653 (1990).

\refis{hier} F.D.M. Haldane, \prl 51, 605, 1983; B.I. Halperin,
\prl 52, 1583, 1983; N. D'Ambrumenil, and R. Morf, \prb 40, 6108, 1989.

\refis{chpap} This point is discussed in detail elsewhere.
See A.H. MacDonald, Helv. Phys. Acta., {\bf 65}, 133 (1992).

\refis{solid}  For a recent summary of experiments which have
been interpreted as providing evidence for a solid state of
electrons see H.W. Jiang, H.L. Stormer, D.C. Tsui,
L.N. Pfeiffer, and K.W. West, Phys. Rev. B {\bf 44}, 8107 (1991).

\refis{theory} See for example K. Esfarjani and S.T. Chui,
Phys. Rev. B {\bf 42}, 10758 (1990) and work cited therein.

\refis{wigner} E.P. Wigner, Phys. Rev. {\bf 46}, 1002 (1934);
Y. E. Losovik, and V.I. Yudson, Pis'ma Zh. Eksp. Teor. Fiz.
{\bf 22}, 26 (1975) [JETP Lett. {\bf 22}, 11 (1975)];
G. Meissner, H. Namaizawa, and M. Voss, \prb 13, 1370, 1976.

\refis{qpi} For recent estimates of quasiparticle interactions see
P. Beran, and R. Morf, \prb 43, 12654, 1991.

\refis{qhe} K. von Klitzing, G. Dorda, and M. Pepper,
Phys. Rev. B {\bf 45}, 494 (1980).

\refis{intro} For an elementary introduction see {\it
A Perspective on the Quantum Hall Effect}, edited by
A.H. MacDonald (Klewer, Boston, 1989).

\refis{fqhe} D.C. Tsui, H.L. St\" ormer, and A.C. Gossard, Phys.
Rev. Lett. {\bf 48}, 1559, (1982).

\refis{laugh} R.B. Laughlin, Phys. Rev. Lett. {\bf 50}, 1395 (1983).

\refis{trug} S.A. Trugman and S.A. Kivelson, Phys. Rev. B {\bf 31}, 5280
(1985).

\endreferences

\centerline{FIGURES}

\item{Fig.[1]} Free energy versus temperature in a fractional
Hall liquid for $\nu$ near $1/m$ with $m=5$.

\item{Fig.[2]} Temperature dependence of the specific heat of the
harmonic Wigner lattice in the strong magnetic field limit.

\item{Fig.[3]} Temperature dependence of the entropy of
harmonic Wigner lattice in the strong magnetic field limit.

\item{Fig.[4]} Temperature dependence of the free energy
of the harmonic Wigner lattice in the strong magnetic field
limit.  The dashed line shows the temperature dependence of the
energy.

\item{Fig.[5]} Free energy difference between liquid and solid
states as a function of filling factor for a series of
temperatures.  The temperautres listed are in units of
$e^2/\ell k_B$.

\item{Fig.[6]} Phase boundary between Wigner solid and
fractional Hall liquid states of the two-dimensional electron
gas at strong magnetic fields near $\nu = 1/5$.

\item{Fig.[7]} Phase boundaries between Wigner solid and
fractional Hall liquid states of the two-dimensional electron
for arbitrarily adjusted incompressible ground state
energies.

\endit